\documentstyle[12pt,a4wide]{article}
\newcommand{\as}{{\alpha_s}}

\newcommand{\GeV}{{\sl\,GeV}}
\newcommand{\real}{{\sl Re\,}}

\newcommand{\Li}{{\rm Li}_2}
\newcommand{\sxi}{{\sqrt\xi}}
\newcommand{\pfrac}[2]{\left(\frac{#1}{#2}\right)}

\begin{document}
\thispagestyle{empty}
\begin{flushright}
MZ-TH/97-39\\
hep-ph/9801255\\
January 1998\\
\end{flushright}
\vspace{0.5cm}
\begin{center}
{\Large\bf On the rigidity of back-to-back top quark pairs}\\[.3cm]
{\Large\bf in \boldmath{$e^+e^-$} annihilation}\\[1.3cm]
{\large S.~Groote, J.G.~K\"orner and J.A.~Leyva\footnote{On leave of absence 
from CIF, Bogot\'a, Colombia}}\\[1cm]
Institut f\"ur Physik, Johannes-Gutenberg-Universit\"at,\\[.2cm]
Staudinger Weg 7, D-55099 Mainz, Germany\\
\end{center}
\vspace{1cm}

\begin{abstract}\noindent
We consider the effect of gluon radiation on the energy of top/antitop 
quarks and on the anticollinearity of top--antitop quark pairs produced in 
$e^+e^-$ annihilation. Our results are presented in terms of the 
$E_q$-dependence of the $t\bar tg$ cross section and the dependence on
the cosine of the opening angle $\theta_{12}$ between top and antitop
for a center of mass energy of $\sqrt{q^2}=500\GeV$. We then go on to
determine mean values for the top quark's energy as well as its longitudinal
and transverse projections, and for the deviation of $\sin\theta_{12}$ and
$\cos\theta_{12}$ from the anticollinearity limits $\sin\theta_{12}=0$ and
$\cos\theta_{12}=-1$. For a center of mass energy of $500\GeV$ we obtain 
$\langle E_q\rangle=248.22\GeV$, $\langle E_L\rangle=247.24\GeV$ and 
$\langle E_T\rangle=4.70\GeV$. Thus, at this energy gluon radiation causes 
a total average energy loss of $0.71\%$ of the top quark's energy. The 
average energy loss in the longitudinal direction is $1.06\%$ and the average 
energy gain in the transverse direction is $1.88\%$. These percentage figures 
go up to $3.77\%$, $5.19\%$ and $6.06\%$, respectively, at $1000\GeV$. For the 
mean of the acollinearity angle $\bar\theta_{12}=180^0-\theta_{12}$ we obtain 
$\langle\bar\theta_{12}\rangle=1.25^0$ at $500\GeV$, the value of which goes
up to $4.62^0$ at $1000\GeV$. From an analysis of the transverse momentum of
the top we find that the mean transverse momentum of the top stays close to 
the mean total momentum of the gluon in the energy range from threshold to 
$1000\GeV$ showing that the gluon momentum has a large mean transverse 
component in this energy range.
\end{abstract}

\newpage

\section{Introduction}
Top quark pairs produced in $e^+e^-$ annihilation provide a very clean and 
kinematically well-defined source of top quarks. To leading order in the 
QCD coupling constant $\as$ the energy of the top quark is exactly given by 
one-half of the center of mass energy. The top and antitop quarks decay so 
fast that there is no visible primary track. Even if the energy of the
top quark is constrained by the beam energy one has to rely on kinematical 
reconstruction from an analysis of the decay products in order to pin 
down the momentum direction of the primary top quark pairs. Since to lowest 
order of QCD the top and antitop are produced exactly back-to-back,
the lowest order anticollinearity provides for a further important 
kinematical constraint on the production kinematics.

An exact knowledge of the momentum configuration of the primary produced 
top quark pairs is important for measurements of the spin observables of 
the top quark. If the top quark's momentum and momentum direction is known,
one can boost to the rest frame of the top quark and do the spin analysis
in the top quark rest system which is the optimal frame from the point
of view of a spin analysis. As concerns spin--spin correlation effects, the 
anticollinearity of the produced top quark pairs is also important for 
optimizing spin--spin correlation measurements~\cite{rigid1}.

Previously it has been emphasized that the bremsstrahlung of gluons off the 
top quark pairs can have important effects on various characteristics of 
the $t\bar t$ final state. For example, the mass determination of the top 
quark and the energy spectrum of secondary leptons will be affected by the 
presence of gluon bremsstrahlung~\cite{rigid2,rigid3,rigid4}. In order to
quantify the effect of gluon bremsstrahlung, the authors of~\cite{rigid5} 
have calculated the gluon energy spectrum in $e^+e^-\rightarrow t\bar tg$ 
as well as the mean energy of the gluon $\langle E_g\rangle$. Having in 
mind the above-mentioned spin analysis we continue in the study of gluon 
emission effects in $e^+e^-\rightarrow t\bar tg$ and determine how the 
energy of the top (or antitop) quark is affected by gluon emission and how 
the anticollinearity of the top--antitop configuration is distorted by 
gluon emission.

Our paper is structured as follows. In Sec.~2 we study the top energy
spectrum and compare it to the bottom energy spectrum. From this we
determine the mean deviation of the top and bottom energy $E_q$ from their 
lowest order value $\frac12\sqrt{q^2}$. A simple analysis shows that the 
mean quark energy $\langle E_q\rangle$ is related to the mean gluon energy 
$\langle E_g\rangle$ calculated in~\cite{rigid5}. Although this result is 
quite plausible at first sight it is nevertheless surprising since the 
calculation of the gluon's mean energy $\langle E_g\rangle$ involves tree 
graph contributions only whereas one needs to bring in loop contributions 
for the determination of the quark's mean energy $\langle E_q\rangle$. In 
Sec.~3 we check on the rigidity of the back-to-back top quark pair 
configuration against gluon radiation by looking at opening angle 
distributions and their mean values. We determine the differential 
$\cos\theta_{12}$-distribution where $\theta_{12}$ is the opening angle of 
the top quark pair and compare to the corresponding distribution in the 
bottom quark case. We then go on to calculate the mean deviations of 
$\cos\theta_{12}$ and $\sin\theta_{12}$ from the anticollinearity limits 
$\cos\theta_{12}=-1$ and $\sin\theta_{12}=0$. In Sec.~4 we study mean 
values of the longitudinal energy $E_L$ and the transverse energy $E_T$ of 
the top quark. The leading contributions to these measures are given by the 
mean energy $\langle E_q\rangle$ calculated in Sec.~2 and by the mean 
opening angle measures $\langle1+\cos\theta_{12}\rangle$ and 
$\langle\sin\theta_{12}\rangle$ calculated in Sec.~4. We also numerically 
evaluate the mean transverse momentum $\langle p_T \rangle$ of the top 
quark relative to the antitop direction and the mean of the opening angle 
itself. Sec.~5 contains our summary and conclusions. In Appendix~A we list 
explicit formulas for the full $O(\as)$ heavy quark production cross section
which enter in the evaluation of the mean values as denominator functions. 
In Appendix~B we list a number of basic rate functions which appear in our 
analytical results. In Appendix~C we present analytical results on first 
moment integrals that are needed for the evaluation of the mean values of 
$\langle E_q\rangle$ and $\langle1+\cos\theta_{12}\rangle$ for scalar and 
pseudoscalar current-induced top pair production.

\section{Quark energy spectrum and $\langle E_q\rangle$}
To begin with we express the differential heavy quark production cross 
section in $e^+e^-\rightarrow q(p_1)\bar q(p_2)g(p_3)$ with 
$q=p_1+p_2+p_3$ in terms of the relevant components of the hadron tensor 
$H^i=(-g^{\mu\nu}+q^\mu q^\nu/q^2)H_{\mu\nu}^i$
\begin{equation}\label{eqn1}
\frac{d\sigma}{dy\,dz}=\frac{\alpha^2}{48\pi q^2}
  \Big(g_{11}H^1(y,z)+g_{12}H^2(y,z)\Big)
\end{equation}
where $y=1-2p_1q/q^2$ and $z=1-2p_2q/q^2$ are fractional energy loss
variables for the heavy quark and antiquark. We shall also use the 
abbreviation $\xi=4m_q^2/q^2$. Another useful variable is the Born term 
velocity of the quark given by $v=\sqrt{1-\xi}$. We closely follow the 
notation employed in~\cite{rigid6}. Thus the superfix $i=1,2$ stands for 
the two current combinations $H^1=\frac12(H^{VV}+H^{AA})$ and 
$H^2=\frac12(H^{VV}-H^{AA})$. In Eq.~(\ref{eqn1}) we have already averaged 
over the relative beam-event orientation. In the terminology 
of~\cite{rigid6} the full hadron tensor components would thus carry the 
additional index $(U+L)$ relevant to the total rate ($U$: unpolarized 
transverse, $L$: longitudinal). However, since we are only concerned with 
the total rate in this paper we have consistently dropped this index. The 
parameters $g_{11}$ and $g_{12}$ specify the electro-weak model dependence 
and are given by
\begin{eqnarray}
g_{11}&=&Q_f^2-2Q_fv_ev_f\real\chi_Z+(v_e^2+a_e^2)(v_f^2+a_f^2)|\chi_Z|^2,
  \nonumber\\
g_{12}&=&Q_f^2-2Q_fv_ev_f\real\chi_Z+(v_e^2+a_e^2)(v_f^2-a_f^2)|\chi_Z|^2,
\end{eqnarray}
where, in the Standard Model,
$\chi_Z(q^2)=gM_Z^2q^2(q^2-M_Z^2+iM_Z\Gamma_Z)^{-1}$, with $M_Z$ and 
$\Gamma_Z$ the mass and width of the $Z^0$ and 
$g=G_F(8\sqrt2\pi\alpha)^{-1}\approx 4.49\cdot 10^{-5}\GeV^{-2}$. $Q_f$ are 
the charges of the final state quarks; $v_e$ and $a_e$, $v_f$ and $a_f$ are 
the electro-weak vector and axial vector coupling constants. For example, in 
the Weinberg-Salam model, one has $v_e=-1+4\sin^2\theta_W$, $a_e=-1$ for 
leptons, $v_f=1-\frac83\sin^2\theta_W$, $a_f=1$ for up-type quarks 
($Q_f=\frac23$), and $v_f=-1+\frac43\sin^2\theta_W$, $a_f=-1$ for down-type 
quarks ($Q_f=-\frac13$). In this paper we use Standard Model couplings
with $\sin^2\theta_W=0.226$.

The hadron tensor components $H^i$ can be obtained from the relevant 
tree-level Feynman diagrams by the above projection
$H^i=(-g^{\mu\nu}+q^\mu q^\nu/q^2)H_{\mu\nu}^i$. They are given by
\begin{eqnarray}
H^1(y,z)&=&N\Bigg[4\xi
  -(4-\xi)\xi\left(\frac1{y^2}+\frac1{z^2}\right)
  -4(4-\xi)\left(\frac1y+\frac1z\right)\nonumber\\&&\qquad\qquad\qquad
  +2(4-\xi)(2-\xi)\frac1{yz}+2(4+\xi)\left(\frac yz+\frac zy\right)\Bigg],
  \label{eqn2}\\
H^2(y,z)&=&N\xi\Bigg[-4
  -3\xi\left(\frac1{y^2}+\frac1{z^2}\right)
  -12\left(\frac1y+\frac1z\right)\nonumber\\&&\qquad\qquad\qquad\qquad
  +6(2-\xi)\frac1{yz}-2\left(\frac yz+\frac zy\right)\Bigg]\label{eqn3}
\end{eqnarray}
where $N=4\pi\as N_CC_F$. The hadron tensor component $H^2(y,z)$ features 
the typical chirality factor $\xi$ which guarantees that $H^{VV}=H^{AA}$ in 
the limit $v\rightarrow 1$ or $m_q\rightarrow 0$.

In order to obtain the energy spectrum of the quark we integrate over 
the fractional energy loss variable $z$ of the antiquark. The integration
limits are given by
\begin{equation}
z_\pm=\frac{y}{4y+\xi}\Big(2-2y-\xi\pm2\sqrt{(1-y)^2-\xi}\Big).
\end{equation}
The integration is straightforward and results in the hadron tensor components
\begin{eqnarray}
H^1(y)&=&2N\Bigg[\Bigg\{\frac{(4-\xi)(2-\xi)}y-2(4-\xi)+(4+\xi)y
  \Bigg\}\ln\pfrac{z_+(y)}{z_-(y)}\nonumber\\&&\qquad
  -2\sqrt{(1-y)^2-\xi}\Bigg\{2\frac{4-\xi}y+\frac{(4-3\xi)y}{4y+\xi}
  -\frac{(16-\xi^2)y}{(4y+\xi)^2}\Bigg\}\Bigg],\\
H^2(y)&=&2N\xi\Bigg[\Bigg\{3\frac{2-\xi}y-6-y\Bigg\}
  \ln\pfrac{z_+(y)}{z_-(y)}\nonumber\\&&\qquad
  -2\sqrt{(1-y)^2-\xi}\Bigg\{\frac6y+\frac{3y}{4y+\xi}
  +\frac{(4-\xi)y}{(4y+\xi)^2}\Bigg\}\Bigg].
\end{eqnarray}

In Fig.~1a we show the resulting $y$-spectra for bottom and top pair
production at $\sqrt{q^2}=500\GeV$ using $m_b=4.83\GeV$, $m_t=175\GeV$ and
a running $\as$ with $\as(m_Z^2)=0.1175$ ($n_f=5$ below $t\bar{t}$-threshold
and $n_f=6$ above $t\bar{t}$-threshold). Both spectra are strongly peaked 
towards low $y$-values due to the infrared (IR) singularity at $y=0$.
The differential rate for bottom pair production is much higher than
that of top pair production because the phase space is much larger for
bottom pair production.
 
Next we determine the mean energy of the quark. Up to $O(\as)$ the mean
energy of the quark is defined by
\begin{equation}\label{eqn4}
\langle E_q\rangle=\frac{\displaystyle\frac12\sqrt{q^2}\sigma({\it Born})
  +\int E_q\frac{d\sigma({\it tree})}{dy\,dz}dy\,dz
  +\frac12\sqrt{q^2}\sigma({\it loop})}{\sigma({\it Born})+\sigma(\as)},
\end{equation}
where $\sigma(\as)$ is the $O(\as)$ cross section
$\sigma(\as)=\sigma({\it tree})+\sigma({\it loop})$. Note that we define
our mean energy relative to the full $O(\as)$ cross section and not to the
Born term cross section as done in~\cite{rigid5}. We shall come back to
this point later on. In order to make the paper self-contained we list
the Born term and $O(\as)$ cross sections appearing in the denominator of
Eq.~(\ref{eqn4}) in Appendix~A. We have checked that the infrared (IR) 
singularity resulting from the tree-graph integration in the numerator of 
Eq.~(\ref{eqn4}) cancels against the IR-singularity of the loop 
contribution, i.e., including the loop contribution, the mean energy of the 
quark is a manifestly IR-safe measure. In fact Eq.~(\ref{eqn4}) can be 
rewritten into a more compact form by using the fractional energy loss 
variable $y$ with $E_q=\frac12\sqrt{q^2}(1-y)$. One has
\begin{equation}\label{eqn5}
\langle y\rangle=\langle 1-2E_q/\sqrt{q^2}\rangle
  =\frac{\displaystyle\int y\frac{d\sigma({\it tree})}{dy\,dz}dy\,dz}
  {\sigma({\it Born})+\sigma(\as)}.
\end{equation}

The calculation of the mean $\langle y\rangle$ is simpler since one only
needs to compute the tree-graph contribution. This is plausible since by
taking the first moment of the differential tree-graph cross section with
regard to $y$ the IR-singular piece in the tree-graph cross section is
cancelled. In the following we shall refer to the mean of variables that
vanish at the IR-point as tree-graph IR-safe measures. For tree-graph
IR-safe measures such as $\langle y\rangle$ the perturbation series starts
at $O(\as)$. At this order the inclusion of the full $O(\as)$ total
cross section in the denominator of Eq.~(\ref{eqn5}) is no longer 
mandatory. By keeping the full $O(\as)$ contribution in the denominator 
one partially includes higher order effects in the evaluation of 
$\langle y\rangle$. In order to be definite, we shall always present our 
results on tree-graph IR-safe measures normalized to the full $O(\as)$ 
cross section.

We give the results of the first moment integration of the 
numerator of Eq.~(\ref{eqn5}) in terms of the first moments $H^{i[y]}$ of 
the two relevant hadron tensor components defined by
\begin{equation}
H^{i[y]}= \int y\,H^i(y,z)dy\,dz.
\end{equation}
We obtain
\begin{equation}
H^{1[y]}=N\Bigg[\frac1{24}(256-224\xi+120\xi^2-8\xi^3-5\xi^4)t_3
  -\frac1{36}(704+336\xi+34\xi^2+15\xi^3)v\Bigg]
\end{equation}
and
\begin{equation}
H^{2[y]}=N\xi\Bigg[\frac1{24}(128-120\xi+36\xi^2+5\xi^3)t_3
  -\frac1{36}(280-118\xi-15\xi^2)v\Bigg].
\end{equation}

Our results for the $y$-moments are simply related to the fractional gluon 
energy $x$-moments $H^{VV[x]}= H^{1[x]}+H^{2[x]}$ and 
$H^{AA[x]}= H^{1[x]}-H^{2[x]}$ calculated in~\cite{rigid5}  
($x=2p_3q/q^2=2E_g/\sqrt{q^2})$ as can be seen by the following reasoning.
Due to $CP$-invariance of the underlying interaction, the $y$-moments
$H^{i[y]}$ for the quarks and the $z$-moments $H^{i[z]}$ for the antiquarks 
are identical, i.e.\ $H^{i[y]}=H^{i[z]}$. From the kinematical relation 
$x=y+z$ one then obtains
\begin{equation}
H^{i[y]}= \frac12H^{i[x]}
\end{equation}
as already noted in~\cite{rigid5}. Our moment results agree with the 
results given in~\cite{rigid5}.

In Fig.~2a we plot the mean value of the fractional energy loss variable
$\langle y\rangle=\langle 1-2E_q/\sqrt{q^2}\rangle$ for bottom and top pair
production as a function of $\sqrt{q^2}$. We first discuss the bottom
pair production case in Fig.~2a where the mean value 
$\langle y \rangle$ shows a sharp rise from threshold and then levels off
at higher energies. To understand the high energy behaviour we consider
the limiting behaviour of the $y$-moments $H^{i[y]}$ as $v\rightarrow1$.
In the high energy limit $v\rightarrow1$ or, equivalently, in the limit 
$m_q\rightarrow0$ we obtain
\begin{equation}\label{eqn6}
H^{1[y]}\rightarrow\frac N3\left[32\ln\pfrac4\xi-\frac{176}3\right],\qquad
H^{2[y]}\rightarrow 0.
\end{equation}
One notes that $H^{1[y]}$ becomes mass singular in this limit due to the 
presence of the quark/antiquark-gluon collinear singularity. There is no 
inconsistency in this singular behaviour from the physics point of view 
because the measure $\langle y\rangle$ is no longer defined in this limit. 
The reason is that the energy of the quark cannot be separately measured in 
the collinear mass zero case. In the corresponding collinear massive 
configuration the energy of the quark can in fact be separately determined by 
e.g.\ a time-of-flight separation. In order to define a physically 
meaningful $\langle y\rangle$ in the $m_q\rightarrow0$ limit one has to 
introduce an angular cut-off. For very small values of the angular cut-off 
one would have to exponentiate the resulting large logarithmic factor taking 
into account multiple gluon emission. Returning to Eq.~(\ref{eqn6}) the 
vanishing of $H^{2[y]}$ is expected since $H^{VV}\rightarrow H^{AA}$ for 
$v\rightarrow 1$ or $m_q\rightarrow 0$ as mentioned before. The limiting 
behaviour of $\langle y\rangle$ following from Eq.~(\ref{eqn6})
can be seen to be given by
\begin{equation}\label{eqn7}
\langle y\rangle\rightarrow\frac\as\pi\ \frac{32\ln(4/\xi)-176/3}
  {36(1+\as/\pi)}.
\end{equation}
The logarithmic growth in Eq.~(\ref{eqn7}) resulting from the mass 
singularity is off-set by the logarithmic fall-off of $\as$.
Looking at Fig.~2a the mean value $\langle y \rangle$ is still slightly
rising at $\sqrt{q^2}=1000\GeV$ showing that the approach of 
$\langle y \rangle$ to its asymptotic value is not very fast.

Turning to the top pair production case in Fig.~2a we first determine the
threshold behaviour of $\langle y \rangle$. In the nonrelativistic limit
$v\rightarrow0$ we obtain
\begin{equation}\label{eqn8}
H^{1[y]}\rightarrow\frac{32}5Nv^5+O(v^7),\qquad
H^{2[y]}\rightarrow\frac{32}5Nv^5+O(v^7).
\end{equation}
The limiting behaviour of the $H^{i[y]}$ can be understood in the following 
way. At threshold the cross section factorizes into the Born term cross 
section $e^+e^-\rightarrow q\bar{q}$ and a gluon production piece. For the 
Born term cross section the $s$-wave vector current contribution
dominates over the $p$-wave axial vector current contribution leading to
$H^{1[y]}\simeq H^{2[y]}$ close to threshold. The leading $v^5$-behaviour 
of the $s$-wave vector current contribution has its origin in the following 
factors: there is one power of $v$ from the flux factor and four powers of 
$v$ from integrating the $y$-weight over gluon phase space. We do not dwell
on the nonleading $O(v^7)$-contributions in Eq.~(\ref{eqn8}) but mention 
that there is some interesting physics concealed in the nonleading 
$O(v^7)$-contributions as discussed in~\cite{rigid5}. Finally, when calculating 
$\langle y\rangle$ one divides by the Born term cross section resulting in 
an overall $v^4$-threshold behaviour. From what has been said about the low
and high energy behaviour of $\langle y \rangle$ it is clear that close to
$t\bar{t}$-threshold the vector current contribution is the dominant 
contribution whereas at high energies the vector and the axial vector 
currents contribute equally to $\langle y \rangle$. The same statement is 
true for the other linear mean values calculated in the following sections. 
We want to mention that the high energy limit ($v\rightarrow 1$) and 
nonrelativistic ($v\rightarrow 0$) limits of $\langle y\rangle$ presented 
in this paper are in complete agreement with the results given 
in~\cite{rigid5}.

The power behaved dependence of $\langle y\rangle$ on $\sqrt{q^2}$ close 
to top--antitop threshold is clearly visible in Fig.~2a. Away from 
threshold $\langle y\rangle$ starts turning over at around $500\GeV$  
in its slow approach to asymptotia. At $500\GeV$ the fractional energy
loss of the top quark amounts to less than 1\% while at $1000\GeV$ the
fractional energy loss is 3.77\% (see Table~\ref{tab1}).
\begin{table}[ht]\begin{center}
\begin{tabular}{|c||c|c|}\hline
\strut mean value&$\sqrt{q^2}=500\GeV$&$\sqrt{q^2}=1000\GeV$\\\hline
$\langle y\rangle$&$0.71\%$&$3.77\%$\\
$\langle 1+\cos\theta_{12}\rangle$&$0.41\%$&$1.99\%$\\
$\langle y(1+\cos\theta_{12})\rangle$&$0.06\%$&$0.57\%$\\
$\langle y_L\rangle$&$1.06\%$&$5.19\%$\\\hline
$\langle\sin\theta_{12}\rangle$&$2.07\%$&$7.30\%$\\
$\langle y\sin\theta_{12}\rangle$&$0.19\%$&$1.23\%$\\
$\langle y_T\rangle$&$1.88\%$&$6.06\%$\\\hline
$\langle x\rangle$&$1.43\%$&$7.53\%$\\
$2\langle p_T\rangle/\sqrt{q^2}$&$1.18\%$&$5.46\%$\\\hline
$\langle\bar\theta_{12}\rangle$&$1.25^0$&$4.62^0$\\\hline
$\xi$&$0.490$&$0.123$\\
$v$&$0.714$&$0.997$\\\hline
\end{tabular}
\caption{\label{tab1}Mean values of different kinematical quantities in 
$e^+e^-\rightarrow t\bar tg$ for $\sqrt{q^2}=500\GeV$ and $1000\GeV$. We 
also list the values of $\xi$ and $v=\sqrt{1-\xi}$.}
\end{center}\end{table}
 
\section{Opening angle distribution \\and the mean values
  $\langle 1+\cos\theta_{12}\rangle$ and $\langle\sin\theta_{12}\rangle$}
Next we determine the differential distribution in the cosine of the 
opening
angle $\cos\theta_{12}$ and the mean deviation from the anticollinearity
limit $\cos\theta_{12}=-1$. The central relation is the relation between 
$\cos\theta_{12}$ and the $(y,z)$-variables which reads
\begin{equation}\label{eqn9}
\cos\theta_{12}=\frac{yz+y+z-1+\xi}{\sqrt{(1-y)^2-\xi}\sqrt{(1-z)^2-\xi}}.
\end{equation}
For later purposes we need to solve Eq.~(\ref{eqn9}) in terms of the 
$y$-variable. One obtains a quadratic equation in $y$ which can be solved 
to give
\begin{eqnarray}\label{eqn10}
y_\pm&=&\frac{(1-z)(1-\cos\theta_{12})+z((2-z)\cos^2\theta_{12}-(z+\xi))}
  {(1+z)^2-\cos^2\theta_{12}((1-z)^2-\xi)}\nonumber\\[7pt]&&
  \pm\frac{\cos\theta_{12}\sqrt{(1-z)^2-\xi}
  \sqrt{4z^2-\xi((1-z)^2-\xi)(1-\cos\theta_{12})}}
  {(1+z)^2-\cos^2\theta_{12}((1-z)^2-\xi)}.
\end{eqnarray}

First we discuss the differential $\cos\theta_{12}$-distribution. With the 
help of relation~(\ref{eqn9}) the differential $(y,z)$-distribution given 
in Eq.~(\ref{eqn1}) is transformed into
\begin{equation}\label{eqn11}
\frac{d\sigma}{d\cos\theta_{12}\,dz}
  =\frac{d\sigma}{dy\,dz}\ \frac{\partial y}{\partial\cos\theta_{12}}
\end{equation}
where the partial derivative is given by
\begin{equation}\label{eqn12}
  \frac{\partial y}{\partial \cos \theta_{12}}
  =\frac{((1-z)^2-\xi)^{1/2}((1-y)^2-\xi)^{3/2}}{(1-y)(2z+\xi)-\xi (z+1)}.
\end{equation}
After substituting for $y$ one can then integrate Eq.~(\ref{eqn11}) with 
regard to $z$. Bearing in mind that the two solutions of the quadratic 
equation Eq.~(\ref{eqn10}) $y_\pm=y_\pm(z,\cos\theta_{12})$ have to be 
substituted in the hadron tensor components $H^i(y,z)$ and the Jacobian 
Eq.~(\ref{eqn12}) it is evident that it is a cumbersome task to do the 
requisite $z$-integration analytically. Instead the integration will be 
done numerically.

A closer look at phase space is of great help in disentangling the
$z$-integration limits and in determining which of the two solutions 
$y_\pm=y_\pm(\cos\theta_{12},z)$ of Eq.~(\ref{eqn10}) have to be 
substituted for the $y$-variable in Eqs.~(\ref{eqn2}), (\ref{eqn3}) 
and~(\ref{eqn12}). In Fig.~3 we show a plot of the $(y,z)$-phase space 
for $m_t=175\GeV$ and $\sqrt{q^2}=500\GeV$. Apart from the phase space 
boundaries defined by $\cos\theta_{12}=\pm1$ we have included the contour 
lines $\cos\theta_{12}=0,\pm0.7$. All contour lines of constant
$\cos\theta_{12}$ must intersect at the points 
$(y_A,z_A)=(1-\sxi,\sxi(1-\sxi)/(2-\sxi))$ and
$(y_B,z_B)=(\sxi(1-\sxi)/(2-\sxi),1-\sxi)$ 
where the antiquark and the quark resp.\ are at rest. The reason is 
that at these two 
points the opening angle $\theta_{12}$ is no longer defined. It is 
evident that the upper limit of the $z$-integration is always given by 
$z_B=1-\sxi$. The lower $z$-limit depends on whether $\cos\theta_{12}$ 
is positive or negative. For positive values of $\cos\theta_{12}$ the lower 
limit is given by $z_A=\sxi(1-\sxi)/(2-\sxi)$. For negative values of 
$\cos\theta_{12}$ the boundary curve $z_C=z_C(\cos\theta_{12})$ can be 
determined by first solving for the minima of the $z=z(\cos\theta_{12},y)$ 
contours in the $(z,y)$-plane and then by inserting the $y$-loci of the
minima into Eq.~(\ref{eqn9}). For the position of the $z$-minima one finds
\begin{equation}\label{eqn13}
y=\frac{z(2-y)}{2z+y}.
\end{equation}
After substituting~(\ref{eqn13}) into Eq.~(\ref{eqn9}) one obtains a 
quartic equation for $z_C=z_C(\cos\theta_{12})$. From the four solutions 
of the quartic equation the one relevant to the case at hand is given by 
\begin{equation}
z_C=\frac{-\xi(1-\cos\theta_{12})+\sqrt{\xi(1-\cos\theta_{12})
  (4(1-\xi)+\xi^2(1-\cos\theta_{12}))}}{4-\xi(1-\cos\theta_{12})}.
\end{equation}
In Fig.~4 we show a plot of the phase space in the 
$(z,\cos\theta_{12})$-plane with boundary curves as determined by the above 
discussion. We have also indicated which of the two solutions of the 
quadratic equation $y_\pm=y(z,\cos\theta_{12})$ of Eq.~(\ref{eqn10}) have to 
be used in the two respective phase space regions $1\le\cos\theta_{12}\le0$
and $0\le\cos\theta_{12}\le-1$.

In Fig.~1b we finally show the differential $\cos\theta_{12}$-distribution
for bottom and top pair production again at the center of mass energy of 
$500\GeV$. As expected the distributions are strongly peaked towards the 
IR-point at $\cos\theta_{12}=-1$. The differential bottom cross section 
dominates over the differential top cross section because of its larger 
phase space.  

Next we determine the mean deviation from the anticollinearity limit 
$\cos\theta_{12}=-1$. To $O(\as)$ the mean of $\cos\theta_{12}$ is 
defined by
\begin{equation}\label{eqn14}
\langle\cos\theta_{12}\rangle=\frac{\displaystyle-\sigma({\it Born})
  +\int\cos\theta_{12}\frac{d\sigma({\it tree})}{dy\,dz}dy\,dz
  -\sigma({\it loop})}{\sigma({\it Born})+\sigma(\as)}.
\end{equation}
Again we have checked by explicit calculation that the IR-singularity from 
the tree graph integration is cancelled by the corresponding IR-singularity 
of the loop contribution. Similar to Eq.~(\ref{eqn5}) one may banish the 
IR-singularity in the tree graph contribution by taking the moment with 
regard to the weight factor $(1+\cos\theta_{12})$. In explicit form one 
adds and subtracts unity in the tree-graph integral, 
$\cos\theta_{12}=1+\cos\theta_{12}-1$. One can then rewrite 
Eq.~(\ref{eqn14}) as
\begin{equation}
\langle 1+\cos\theta_{12}\rangle=\frac{\displaystyle\int(1+\cos\theta_{12})
  \frac{d\sigma({\it tree})}{dy\,dz}dy\,dz}
  {\sigma({\it Born})+\sigma(\as)}.
\end{equation}
What is needed are the first moment integrals $H^{i[t]}$ taken with regard 
to the moment variable $t=1+\cos\theta_{12}$ for the numerator functions. 
Since $\langle 1+\cos\theta_{12}\rangle$ is a tree-graph IR-safe measure on 
needs to consider only the tree-graph components of the $O(\as)$ hadron 
tensor as explicated after Eq.~(\ref{eqn5}). The requisite moment 
integrations can be done analytically. In fact similar integrations appear 
in the calculation of longitudinal spin--spin 
correlations~\cite{rigid8,rigid9}, and some of the analytical results can 
be taken from these papers. One obtains
\begin{eqnarray}
H^{1[t]}&=&N\Bigg[8+16\sxi-62\xi+56\xi\sxi+3\xi^2+3\xi^2\sxi
  +4(4-5\xi-5\xi^2)\frac1v\nonumber\\&&
  +\left(128-128\xi-12\xi^2+39\xi^3-3\xi^4-32(4-\xi)v^3\right)
  \frac{t_3}{4v^2}\nonumber\\&&
  +2(4-\xi)(2-\xi)(t_9-t_{16})
  -(64-48\xi-116\xi^2+31\xi^3-3\xi^4)\frac{t_{13}}{2v^2}\nonumber\\&&
  -4(4+\xi)(1-2\xi)t_{14}+2(16-40\xi+31\xi^2-10\xi^3)\frac{t_{15}}{v^3}
  \Bigg],\label{eqn15}\\
H^{2[t]}&=&N\xi\Bigg[98-68\sxi-3\xi-3\xi\sxi-4(7-\xi)\frac1v
  +3(32-25\xi^2+\xi^3-96v^3)\frac{t_3}{4v^2}\nonumber\\&&
  +6(2-\xi)(t_9-t_{16})+(208-168\xi+35\xi^2-3\xi^3)\frac{t_{13}}{2v^2}
  \nonumber\\&&
  -4(13+3\xi)t_{14}+6(4-13\xi+8\xi^2)\frac{t_{15}}{v^3}\Bigg].\label{eqn16}
\end{eqnarray}
The rate functions $t_i$ (for $i=3,9,13,14,15,16$) appearing in 
Eqs.~(\ref{eqn15}) and~(\ref{eqn16}) are listed in Appendix~B. As will be
discussed in Sec.~4, $\langle 1+\cos\theta_{12}\rangle$ together 
with $\langle y\rangle$ determine the leading contribution to the mean of 
the longitudinal energy $\langle E_L\rangle$.

In Fig.~2b we show a plot of the $\sqrt{q^2}$-dependence of
$\langle 1+\cos\theta_{12}\rangle$ for bottom and top production. In the
bottom case $\langle 1+\cos\theta_{12}\rangle$ shows a quick rise from
threshold and very quickly reaches its asymptotic value away from threshold.
The asymptotic value is determined by the limiting behaviour of the 
$t$-moments in the high energy limit $v\rightarrow1$. These are given 
by
\begin{equation}
H^{1[t]}\rightarrow 24N,\qquad H^{2[t]}\rightarrow0,
\end{equation}
which translates into
\begin{equation}\label{eqn17}
\langle 1+\cos\theta_{12}\rangle\rightarrow\frac\as\pi\ \frac2{1+\as/\pi}.
\end{equation}
In Fig.~2b we show a plot of the energy dependence of
$\langle 1+\cos\theta_{12}\rangle$ for bottom pair production. After a 
steep rise from threshold the mean value $\langle 1+\cos\theta_{12}\rangle$
starts to decrease again at around $200\GeV$. The logarithmic fall-off
which sets in beyond $200\GeV$ is due to the combined effect of
$\langle 1+\cos\theta_{12}\rangle$ reaching its asymptotic value and the
logarithmic running of the coupling constant $\as$. 

Turning to top pair production in Fig.~2b we first discuss the threshold 
behaviour of $\langle 1+\cos\theta_{12}\rangle$. In the nonrelativistic 
limit $v\rightarrow 0$  the threshold dependence of the moments $H^{i[t]}$ 
is given by
\begin{equation}
H^{1[t]}\rightarrow\frac{256}{75}Nv^5+O(v^7),\qquad
  H^{2[t]}\rightarrow\frac{256}{75}Nv^5+O(v^7).
\end{equation}
One thus again has a $v^4$-threshold behaviour for the mean 
$\langle 1+\cos\theta_{12}\rangle$. The $v^4$-threshold behaviour can 
clearly be discerned in Fig.~2b. At around $600\GeV$ 
$\langle 1+\cos\theta_{12}\rangle$ starts turning over in 
its approach to its asymptotic value. Numerical values of
$\langle 1+\cos\theta_{12}\rangle$ for top pair production are given in 
Table~\ref{tab1}.

Next we determine the mean value of $\sin\theta_{12}$ which, according to the 
discussion in Sec.~4, gives the leading contribution to the mean transverse 
energy of the quark. The mean value $\langle\sin\theta_{12}\rangle$ 
provides another measure of how much gluon radiation distorts the lowest 
order back-to-back configuration of the quark--antiquark pair. First we 
express $\sin\theta_{12}$ in terms of the $(y.z)$-phase space variables.
One has
\begin{equation}
\sin\theta_{12}=\frac{\sqrt{4yz(1-y-z)-\xi(y+z)^2}}
{\sqrt{(1-y)^2-\xi}\sqrt{(1-z)^2-\xi}}.
\end{equation}
Similar to $\langle1+\cos\theta_{12}\rangle$ the mean value of 
$\sin\theta_{12}$ is determined in terms of the $O(\as)$ tree graph 
contribution alone since $\sin\theta_{12}$ vanishes at the IR-singular 
point. In the terminology of the previous sections 
$\langle\sin\theta_{12}\rangle$ is a tree-graph IR-safe measure. The 
necessary integrations have been done numerically. The top production 
results are shown in Fig.~5b where we plot $\langle\sin\theta_{12}\rangle$ 
as a function of the center of mass energy $\sqrt{q^2}$. A comparison with 
Fig.~5a shows that $\langle\sin\theta_{12}\rangle$ is smaller than 
$\langle 1+\cos\theta_{12}\rangle$ over the whole range from top--antitop 
threshold to $1000\GeV$. Qualitatively this is not difficult to understand 
since close to the IR-point $\cos\theta_{12}=-1$ where the cross section 
is largest the weight function $\sin\theta_{12}=\sqrt{1-\cos^2\theta_{12}}$ 
clearly dominates over the weight $\cos\theta_{12}$ when considered as 
functions of $\cos\theta_{12}$. In fact $\sin\theta_{12}$ has an infinite 
slope at the IR-point compared to the unit slope of $\cos\theta_{12}$. One 
therefore expects 
$\langle 1+\cos\theta_{12}\rangle<\langle\sin\theta_{12}\rangle$
which is quantitatively borne out as Figs.~5a and~5b show. Numerical values for
$\langle\sin\theta_{12}\rangle$ at the two center of mass energies $500$ 
and $1000\GeV$ are listed in Table~\ref{tab1}. 

Of interest is also the mean value of the opening angle $\theta_{12}$ or 
its complement, the acollinearity angle $\bar\theta_{12}=180^0-\theta_{12}$. 
In Fig.~6 we show a plot of the energy dependence of the mean of the 
acollinearity angle $\langle\bar\theta_{12}\rangle$. 
$\langle\bar\theta_{12}\rangle$ rises smoothly from threshold to a value
of $1.25^0$ at $500\GeV$ and further to $4.62^0$ at $1000\GeV$. A good 
estimate of $\langle\bar\theta_{12}\rangle$ can be obtained by inverting 
the corresponding numerical values for $\langle\sin\theta_{12}\rangle$ in 
Table~\ref{tab1} which gives $\arcsin(\langle\sin\theta_{12}\rangle)=1.19^0$ 
and $4.19^0$, respectively. Using instead $\langle 1+\cos\theta_{12}\rangle$ 
for the corresponding estimate one obtains $5.16^0$ and $11.45^0$ which 
considerably overestimates the true values of $\langle\bar\theta_{12}\rangle$. 
The reason that $\langle\sin\theta_{12}\rangle$ gives a good estimate of the 
mean of the acollinearity angle $\langle\bar\theta_{12}\rangle$ can again 
be traced to the fact that the weight factor $\sin\theta_{12}$ (considered 
as function of $\cos\theta_{12}$) has a strong support close to the IR-point 
$\cos\theta_{12}=-1$.

\section{Mean values of the transverse \\and longitudinal energy}
Up to now we have been separately considering the energy loss of the quark 
and the deviation of the acollinearity angle $\theta_{12}$ from the
anticollinearity limit $\theta_{12}=180^0$ due to gluon radiation. From the 
physics point of view it is also interesting to consider the energy loss of 
the quark projected in the longitudinal direction as well as the build-up 
of transverse energy resulting from gluon radiation. Here the longitudinal 
and transverse energies are defined relative to the momentum direction of 
the antiquark and are thus defined by $E_L=-E_q\cos\theta_{12}$ and 
$E_T=E_q\sin\theta_{12}$. Again we shall calculate mean values of these two 
quantities as a measure of the importance of gluon radiation. As before we 
rewrite $\langle E_L\rangle$ and $\langle E_T\rangle$ in terms of 
tree-graph IR-safe measures. For the mean longitudinal energy one has
\begin{equation}\label{eqn18}
\langle E_L\rangle=\frac12\sqrt{q^2}\Big[1-\langle 1+\cos\theta_{12}\rangle
  -\langle y\rangle+\langle y(1+\cos\theta_{12})\rangle\Big]
\end{equation}
or, written in terms of the fractional longitudinal energy loss variable
$y_L=1-2E_L/\sqrt{q^2}$, one has
\begin{equation}\label{eqn19}
\langle y_L\rangle=\Big[\langle 1+\cos\theta_{12}\rangle
+\langle y\rangle-\langle y(1+\cos\theta_{12})\rangle\Big].
\end{equation}
The mean values appearing in Eq.~(\ref{eqn19}) must satisfy three 
geometrical inequalities which follow from the geometrical inequalities
$y\le y_L$, $(1+\cos\theta_{12})\le y_L$ and $y-y(1+\cos\theta_{12})\le y_L$. 
Since these inequalities are true on an event-by-event basis they must 
also hold for their means. One can check that the relevant numerical 
entries in Table~\ref{tab1} satisfy the three inequalities. On top of the 
geometrical inequalities one may classify the mean values of the different 
weight variables in Eq.~(\ref{eqn19}) in terms of powers of smallness since 
the differential cross section is strongly peaked towards the IR-region. Thus 
one expects that the leading contributions to $\langle E_L\rangle$ come 
from the linear terms $\langle 1+\cos\theta_{12}\rangle$ and 
$\langle y\rangle$ whereas the quadratically small term 
$\langle y(1+\cos\theta_{12})\rangle$ is expected to be a nonleading 
effect. This qualitative picture is borne out by the numerical values in 
Table~\ref{tab1} and by the curves in Fig.~5a where the energy dependence 
of the above mean values are combined in one common plot.

The mean of the fractional transverse energy $y_T=2E_T/\sqrt{q^2}$ can be
written as
\begin{equation}\label{eqn20}
\langle y_T\rangle=\Big[\langle\sin\theta_{12}\rangle
  -\langle y\sin\theta_{12}\rangle\Big].
\end{equation}
One now has the geometrical inequality
$\langle y_T\rangle\le\langle\sin\theta_{12}\rangle$ which can be seen
to be satisfied by the relevant numerical entries in Table~\ref{tab1}. 
The leading contribution is expected to arise from the linear term 
$\langle\sin\theta_{12}\rangle$ whereas the contribution of the term 
$\langle y\sin\theta_{12}\rangle$ is nonleading. The numerical values in 
Table~\ref{tab1} as well as the relevant curves in Fig.~5b confirm this 
picture.

Even though we do not present numerical results for the bottom production
case we list the high energy limits of the mean values 
$\langle y(1+\cos\theta_{12})\rangle$, $\langle\sin\theta_{12}\rangle$ and 
$\langle y\sin\theta_{12}\rangle$ appearing in Eqs.~(\ref{eqn19}) 
and~(\ref{eqn20}) for the sake of completenes. Including also the result 
Eq.~(\ref{eqn17}) one has
\begin{eqnarray}
\langle 1+\cos\theta_{12}\rangle&\rightarrow&
  \frac\as\pi\ \frac2{1+\as/\pi},\nonumber\\
\langle y(1+\cos\theta_{12})\rangle&\rightarrow&
  \frac\as\pi\ \frac{20}{27(1+\as/\pi)},\nonumber\\
\langle\sin\theta_{12}\rangle&\rightarrow&
  \frac\as\pi\ \frac{10\pi}{9(1+\as/\pi)},\nonumber\\
\langle y\sin\theta_{12}\rangle&\rightarrow&
  \frac\as\pi\ \frac\pi{2(1+\as/\pi)}.\label{eqn21}
\end{eqnarray}
Where applicable the limiting values can be seen to satisfy the geometrical 
inequalities written down before. The limiting values show the same 
hierarchy in terms of powers of smallness as discussed before for top 
production below $1000\GeV$. However, the hierarchy is not as pronounced as 
in the top production case below $1000\GeV$. As concerns the limiting 
values in Eqs.~(\ref{eqn21}) we want to add two remarks. In the mass zero 
limit treated above the transverse energy of the quark is equal to its 
transverse momentum. Also the mean of the longitudinal energy 
$\langle E_L\rangle$ and the longitudinal fractional energy loss 
$\langle y_L\rangle$ in Eqs.~(\ref{eqn18}) and~(\ref{eqn19}) are mass 
singular as manifested by the contribution of $\langle y \rangle$ to 
$\langle E_L\rangle$ and $\langle y_L\rangle$ which by itself is mass 
singular (see Sec.~2).

As Table~\ref{tab1} shows, gluon radiation causes an average energy loss 
of 1.06\% of the top quark's energy in the longitudinal direction and an 
average energy gain of 1.88\% in the transverse direction at this energy. 
At $1000\GeV$ the corresponding percentage figures are 5.19\% and 6.06\%.
This has to be compared with the total energy loss of $0.71\%$ at $500\GeV$ 
and $3.77\%$ at $1000\GeV$ as calculated in Sec.~2. The gain in transverse 
energy is larger than the loss of longitudinal energy in this energy range 
where this effect becomes smaller as the center of mass energy increases. 
Qualitatively this behaviour may be understood from what is referred to as 
the ``dead-cone'' effect of gluon emission from a heavy quark~\cite{rigid10}. 
Gluon emission is suppressed in an angular cone of size 
$\theta\simeq\sqrt\xi= 2m_q/\sqrt{q^2}$ around the quark--antiquark direction, 
i.e.\ the emitted gluon has a strong momentum component transverse to the 
quark--antiquark direction. Because of momentum conservation the transverse 
momentum of the gluon has to be balanced by the transverse momentum of the 
quark. A good estimate of the fractional  transverse momentum of the heavy 
quark in the energy region under consideration can be obtained by rescaling 
$y_T$ by the factor $v=p/E=2p/\sqrt{q^2}$ where $p$ is the lowest order 
momentum of the quark (see Table~\ref{tab1}). Taking the rescaled values of 
$\langle y_T\rangle$ and comparing it to the mean fractional total energy 
of the gluon $\langle x\rangle$ in Table~\ref{tab1} one concludes from 
momentum conservation that, on average, the gluon must have a large 
transverse momentum or energy component. By comparing the relevant numbers 
at $500$ and $1000\GeV$ one notes that this effect becomes smaller as the 
energy is raised in agreement with the shrinking of the ``dead-cone''
predicted in~\cite{rigid10}.

In order to get a more quantitative handle on the ``dead-cone'' effect
we show a plot of the $q^2$-dependence of the mean fractional gluon energy 
(or momentum) $\langle 2E_g/\sqrt{q^2}\rangle=\langle x\rangle$ and the mean 
fractional transverse momentum $\langle 2p_T/\sqrt{q^2}\rangle$ of the
top in Fig.~7. In the whole range from threshold up to $1000\GeV$ the mean 
transverse momentum of the top quark is almost balanced by the mean total 
momentum of the gluon where the approximate equality of the two means 
degrades as the energy increases. This proofs that the emitted gluon has a 
dominant transverse momentum component, and that the preference for 
transverse emission is degraded as the energy increases, just as expected 
from the ``dead-cone'' effect.

\section{Summary and conclusions}
We have discussed in some detail how gluon radiation affects the lowest 
order back-to-back configuration of pair produced top quarks in  $e^+e^-$ 
annihilation. We have introduced various measures that serve to quantify 
gluon radiation effects in this reaction and we have studied their 
energy dependence. These measures are related to the overall energy loss 
and the energy loss in the longitudinal direction, or, in the case of the 
transverse energy, to the energy gain in the transverse direction. We 
have also introduced two angular measures related to the opening angle 
$\theta_{12}$ between quark and antiquark.

When possible we have given analytical expressions for the energy and
momentum measures $\langle E_q\rangle$, $\langle E_T\rangle$, 
$\langle E_L\rangle$ and $\langle p_T\rangle$ and for the angular measures 
$\langle 1+\cos\theta_{12}\rangle$ and $\langle\sin\theta_{12}\rangle$. We 
have provided numerical results for all these measures in the energy range 
from close to top--antitop threshold up to $1000\GeV$. For some of the
measures we have compared the top--antitop results to the corresponding
figures for bottom pair production including a discussion of the high 
energy limit $v\rightarrow 1$ limits of the various measures. The 
numerical predictions and their energy dependence can directly be compared to 
experimental data on top pair production at the proposed future linear 
colliders. The values of these measures also serve to quantify the effect 
of gluon radiation on the lowest order back-to-back configuration of the 
produced top quark pairs.

Compared to the lowest order energy of the top $E_q=\frac12\sqrt{q^2}$ the
mean loss of longitudinal energy is smaller than the mean gain in transverse
energy in the range of c.m.\ energies from threshold to $1000\GeV$. At 
$500\GeV$ the mean transverse energy exceeds the mean longitudinal energy 
loss by about a factor of two. In this sense the top quark is more rigid in 
the longitudinal direction than in the transverse direction. This behaviour 
can be traced to an effect which is referred to as the ``dead-cone'' effect 
in the literature. In massive quark pair production with additional gluon 
radiation the gluon tends to be emitted away from the heavy quark and 
antiquark directions. Since the mean momentum of the gluon is comparatively
large ($\langle x \rangle=2\langle y \rangle$) and since the transverse
momentum of the gluon has to be 
balanced by the transverse motion of the quark (or antiquark) one obtains 
comparatively large values for the mean transverse energy and momentum of
the quark (or antiquark). 
   
\section*{Appendix A: $O(\as)$ cross section}
\setcounter{equation}{0}\def\theequation{A\arabic{equation}}
The complete $O(\as)$ cross section for heavy quark production in
$e^+e^-\rightarrow q\bar qg$ has been calculated before
in~\cite{rigid6,rigid7}. In order to make the paper self-contained we list 
the results in this Appendix. For the Born term cross section 
$\sigma({\it Born})$ one obtains
\begin{eqnarray}
\sigma({\it Born})=
  \Big(g_{11}\sigma^1({\it Born})+g_{12}\sigma^2({\it Born})\Big)
\end{eqnarray}
where
\begin{equation}
\sigma^1({\it Born})=\frac{\pi\alpha^2vN_C}{3q^2}(4-\xi),\qquad
  \sigma^2({\it Born})=\frac{\pi\alpha^2vN_C}{3q^2}3\xi.
\end{equation}
The $O(\as)$ tree- and loop-contributions are given in terms of the hadron 
tensor components $H^i$ defined in Eq.~(\ref{eqn1}). One has
\begin{eqnarray}
H^1(\as)&=&N\Big[\frac32(4-\xi)(2-\xi)v
  +\frac14(192-104\xi-4\xi^2+3\xi^3)t_3\nonumber\\&&
  -2(4-\xi)(2-\xi)(t_8-t_9)-4(4-\xi)v(t_{10}+2t_{12})\Big],\\
H^2(\as)&=&N\xi\Big[\frac32(18-\xi)v+\frac34(24-8\xi-\xi^2)t_3\nonumber\\&&
  -6(2-\xi)(t_8-t_9)-12v(t_{10}+2t_{12})\Big].
\end{eqnarray}
The rate functions $t_i$ (for $i=3,8,9,10,12$) are given in Appendix~B.

\section*{Appendix B: Rate integrals and first moment integrals}
\setcounter{equation}{0}\def\theequation{B\arabic{equation}}
The analytical expressions of the $O(\as)$ cross section and the mean value 
$\langle 1+\cos\theta_{12}\rangle$ involve a number of basic rate functions 
which are collected in this Appendix.
\begin{eqnarray}
t_3&=&\ln\left(\frac{1+v}{1-v}\right),\nonumber\\
t_8&=&\ln\left(\frac\xi 4\right)\ln\left(\frac{1+v}{1-v}\right)
  +\Li\left(\frac{2v}{1+v}\right)-\Li\left(-\frac{2v}{1-v}\right)-\pi^2,
  \nonumber\\
t_9&=&2\ln\left(\frac{2(1-\xi)}\sxi\right)
  \ln\left(\frac{1+v}{1-v}\right)
  +2\left(\Li\left(\frac{1+v}2\right)-\Li\left(\frac{1-v}2\right)\right)
  \,+\nonumber\\&&
  +3\left(\Li\left(-\frac{2v}{1-v}\right)
  -\Li\left(\frac{2v}{1+v}\right)\right),\nonumber\\
t_{10}&=&\ln\left(\frac4\xi\right),\quad
t_{12}\ =\ \ln\left(\frac{4(1-\xi)}\xi\right),\quad
t_{13}\ =\ \ln\pfrac{1+v}{2-\sxi},\nonumber\\
t_{14}&=&\ln\pfrac4\xi\ln\pfrac{1+v}{2-\sxi}\nonumber\\&&
  +2\Li\pfrac{2-\sxi}2-2\Li\pfrac\sxi2+\Li\pfrac{1-v}2-\Li\pfrac{1+v}2,
  \nonumber\\
t_{15}&=&\left(\ln\pfrac{1+v}{1-v}+\ln\pfrac\sxi{2-\sxi}\right)^2\nonumber\\&&
  -4\Li\left(\sqrt{\frac{1-v}{1+v}}\right)
  +2\Li\pfrac{2-\sxi}{1+v}+2\Li\pfrac{1-v}{2-\sxi},\nonumber\\
t_{16}&=&\ln\pfrac{1+v}{1-v}\ln\pfrac{4v^4}{\xi(1+v)^2}
  -\Li\pfrac{2v}{(1+v)^2}+\Li\pfrac{-2v}{(1-v)^2}\nonumber\\&&\qquad\qquad
  +\frac12\Li\left(-\frac{(1-v)^2}{(1+v)^2}\right)
  -\frac12\Li\left(-\frac{(1+v)^2}{(1-v)^2}\right).
\end{eqnarray}

\section*{Appendix C: Scalar and pseudoscalar currents}
\setcounter{equation}{0}\def\theequation{C\arabic{equation}}
In this Appendix we present expressions for the hadron tensor components and 
their first moments which allows one to calculate the mean values 
$\langle y\rangle$ and $\langle 1+\cos\theta_{12}\rangle$ for the scalar 
$(S)$ and pseudoscalar $(P)$ current contributions 
$H^1=\frac12(H^{SS}+H^{PP})$ and $H^2=\frac12(H^{SS}-H^{PP})$.
These expressions would be needed for the assessment of gluon emission 
effects in the heavy quark production process $e^+e^-\rightarrow q\bar q(g)$ 
mediated by scalar $(S)$ or pseudoscalar $(P)$ particles as resulting from 
e.g.\ Higgs exchange. One has
\begin{eqnarray}
H^{1[y]}&=&N\xi\Big[\frac32(2-\xi)v-\frac14(8-8\xi+3\xi^2)t_3\Big],\\
H^{2[y]}&=&N\Big[\frac16(2-\xi)(16-16\xi+7\xi^2)t_3
  -\frac19(76-76\xi+21\xi^2)v\Big],\\
H^{1[t]}&=&N\xi\Big[\frac{8\xi}v-8(2-\sxi)
  -2(4-\xi-2\xi^2-4v^3)\frac{t_3}{v^2}-2(2-\xi)(t_9-t_{16})\nonumber\\&&
  -4(4-\xi)\frac{t_{13}}{v^2}+2(6+\xi)t_{14}
  -2(4-12\xi+7\xi^2)\frac{t_{15}}{v^3}\Big],\\
H^{2[t]}&=&N\Big[8(7-9\xi+\xi^2)\frac1v-2(34-36\sxi+5\xi-7\xi\sxi)\nonumber\\&&
  +(32-48\xi+15\xi^2+5\xi^3-16(2-\xi)v^3)\frac{t_3}{2v^2}\nonumber\\&&
  +2(2-\xi)^2(t_9-t_{16})-(80-120\xi+31\xi^2-3\xi^3)\frac{t_{13}}{v^2}
  \nonumber\\&&
  +2(4+10\xi+\xi^2)t_{14}+2(8-22\xi+20\xi^2-7\xi^3)\frac{t_{15}}{v^3}\Big].
\end{eqnarray}

\newpage

\vspace{1cm}
\centerline{\Large\bf Figure Captions}
\vspace{.5cm}
\newcounter{fig}
\begin{list}{\bf\rm Fig.\ \arabic{fig}:}{\usecounter{fig}
\labelwidth1.6cm\leftmargin2.5cm\labelsep.4cm\itemsep0ex plus.2ex}
\item Differential (a) $y$- and (b) $\cos\theta_{12}$-distribution for 
  $e^+e^-\rightarrow b\bar bg$ and $e^+e^-\rightarrow t\bar tg$ at 
  $\sqrt{q^2}=500\GeV$ ($m_t=175\GeV$, $m_b=4.83\GeV$)
\item Mean values of (a) the fractional energy loss variable $y$ of the
  bottom and top quark and (b) the variable $t=1+\cos\theta_{12}$ as 
  functions of $\sqrt{q^2}$ for bottom and top quark pair production 
  ($m_t=175\GeV$, $m_b=4.83\GeV$)
\item $(y,z)$-phase space for $e^+e^-\rightarrow t\bar tg$ ($m_t=175\GeV$, 
  $\sqrt{q^2}=500 \GeV$). Also shown are contour lines for constant values of 
  $\cos\theta_{12}$. To the right and to the left of the zero contour line 
  $\cos\theta_{12}=0$ are the phase space regions $0\le\cos\theta_{12}\le 1$
  and $-1\le\cos\theta_{12}\le 0$. All contour lines of constant 
  $\cos\theta_{12}$ intersect at points $A$ and $B$. Point $C$ denotes the 
  minimum of the $\cos\theta_{12}$ contour lines for negative values of 
  $\cos\theta_{12}$. The IR-point is at the origin.
\item $(z,\cos\theta_{12})$-phase space for $e^+e^-\rightarrow t\bar tg$ 
  ($m_t=175\GeV$, $\sqrt{q^2}=500\GeV$). The IR-point is at the lower left 
  corner. The dark-shaded $y_+$- and the light-shaded $y_-$-regions 
  correspond to the two solutions of the quadratic equation 
  $y_\pm=y_\pm(z,\cos\theta_{12})$
\item Comparison of the mean values of (a) $y_L=1-2E_L/\sqrt{q^2}$, $y$, 
  $1+\cos\theta_{12}$, and $y(1+\cos\theta_{12})$ and (b) 
  $y_T=2E_T/\sqrt{q^2}$, $\sin\theta_{12}$, $y$, and $y\sin\theta_{12}$ for 
  top pair production in $e^+e^-\rightarrow t\bar tg$ as function of 
  $\sqrt{q^2}$
\item Mean value of the acollinearity angle $\bar{\theta}_{12}$ 
  in $e^+e^-\rightarrow t\bar tg$ as function of $\sqrt{q^2}$
\item Mean values of the fractional gluon energy $x$ and fractional 
  transverse momentum $2p_T/\sqrt{q^2}$ of the top quark in 
  $e^+e^-\rightarrow t\bar tg$ as functions of $\sqrt{q^2}$
\end{list}
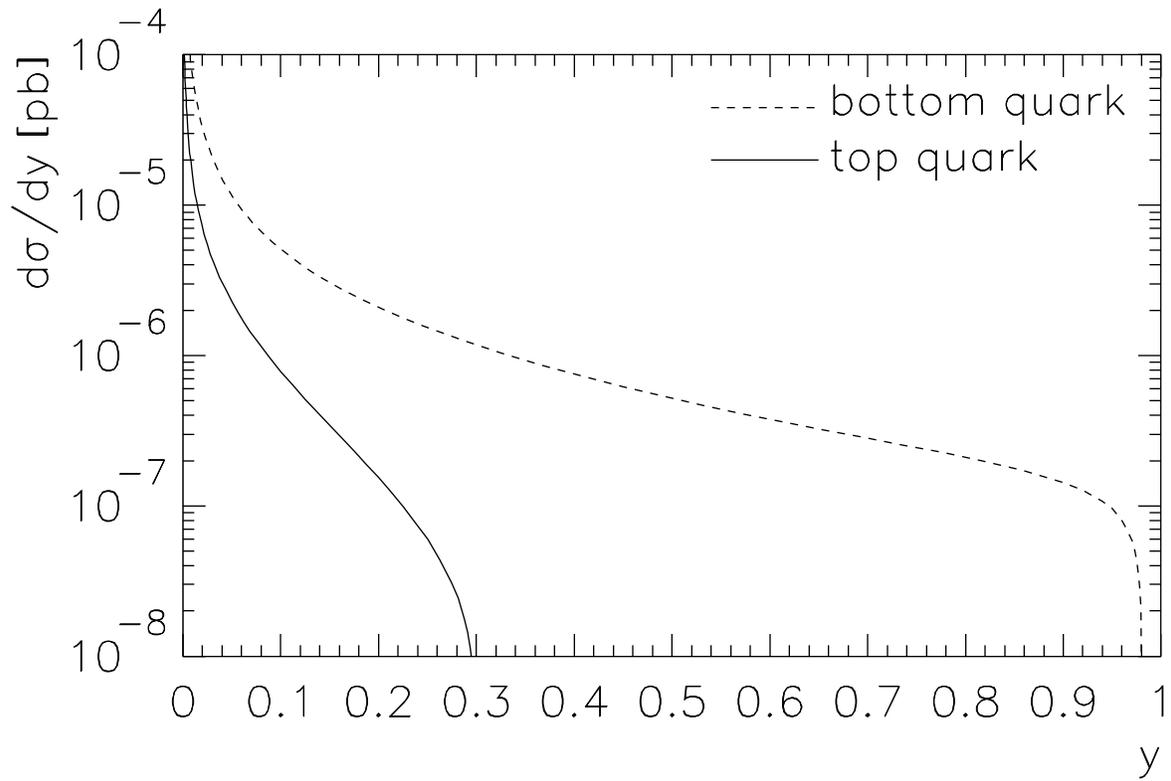
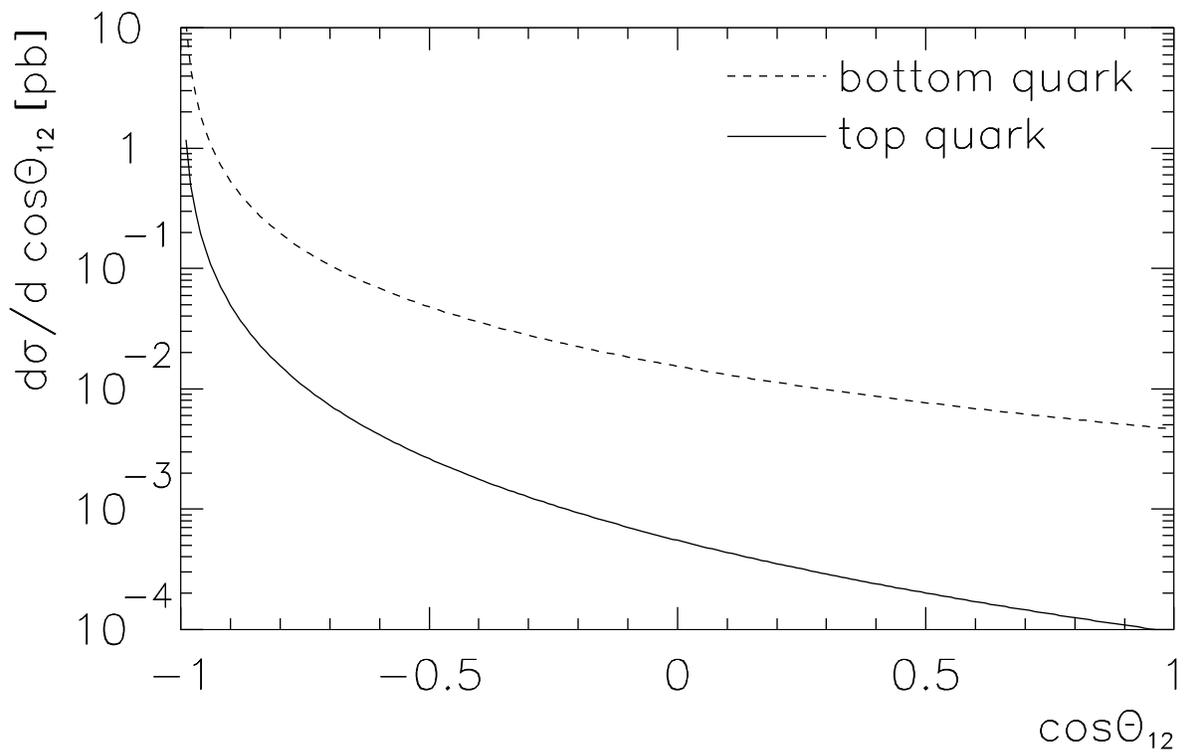
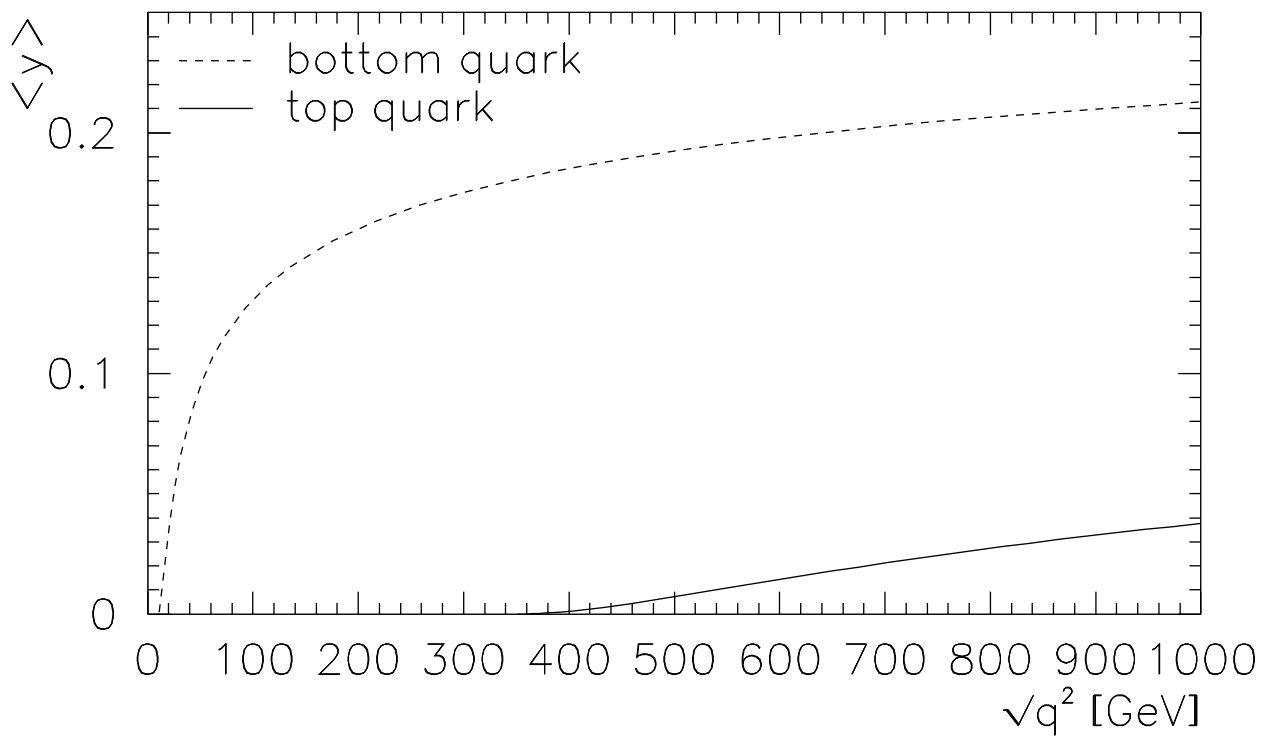
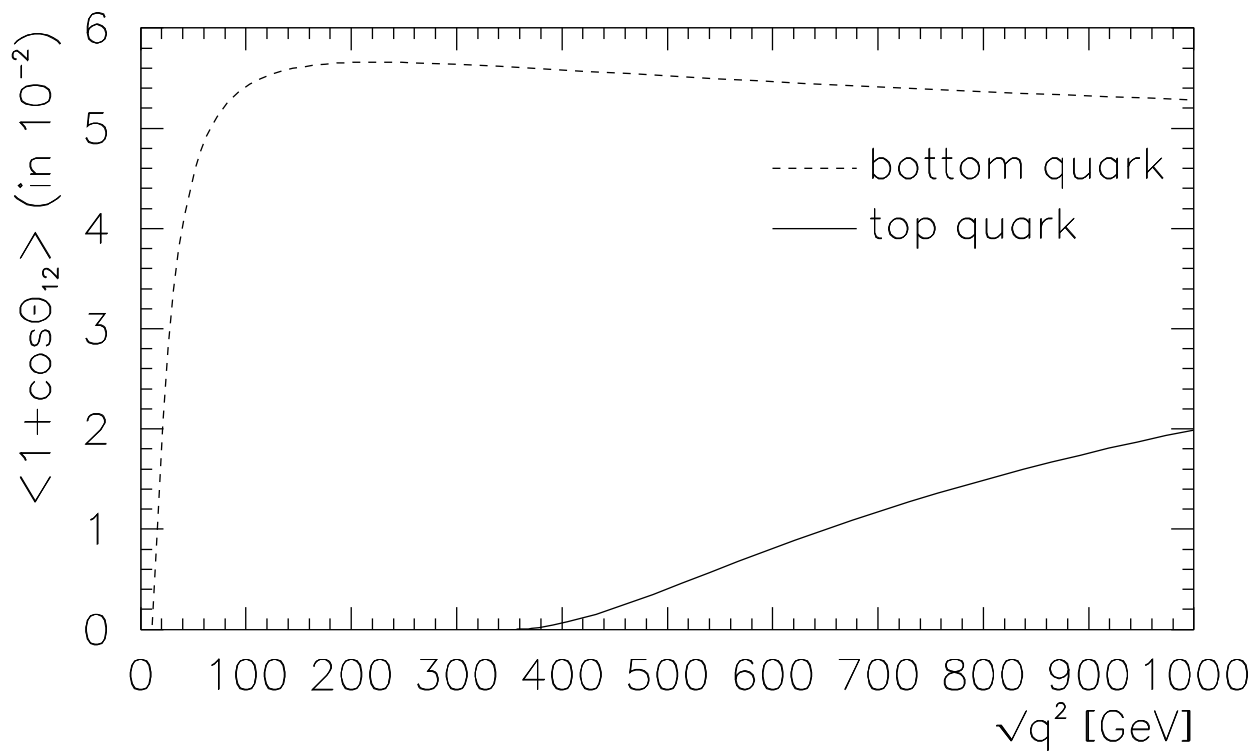
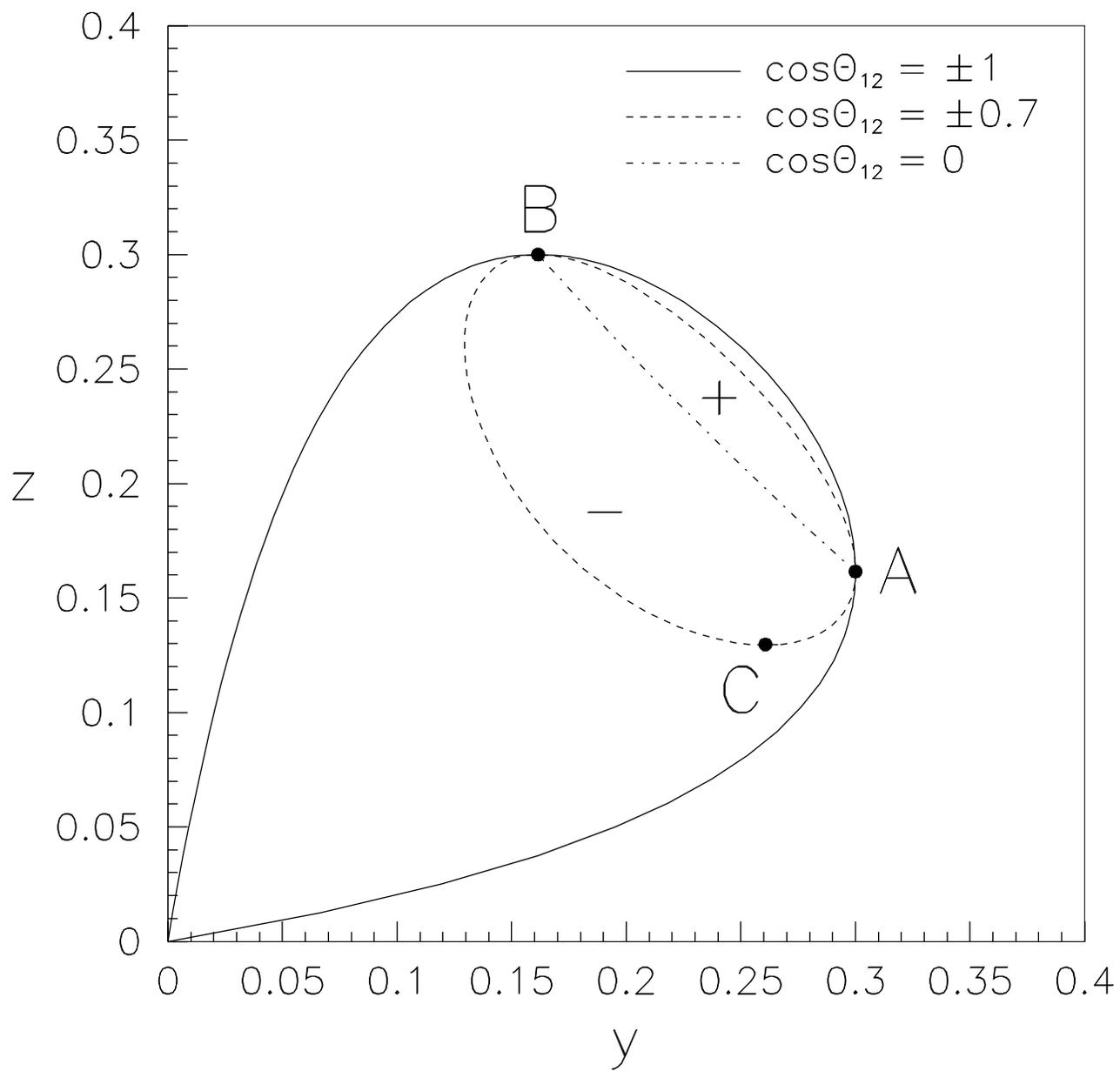
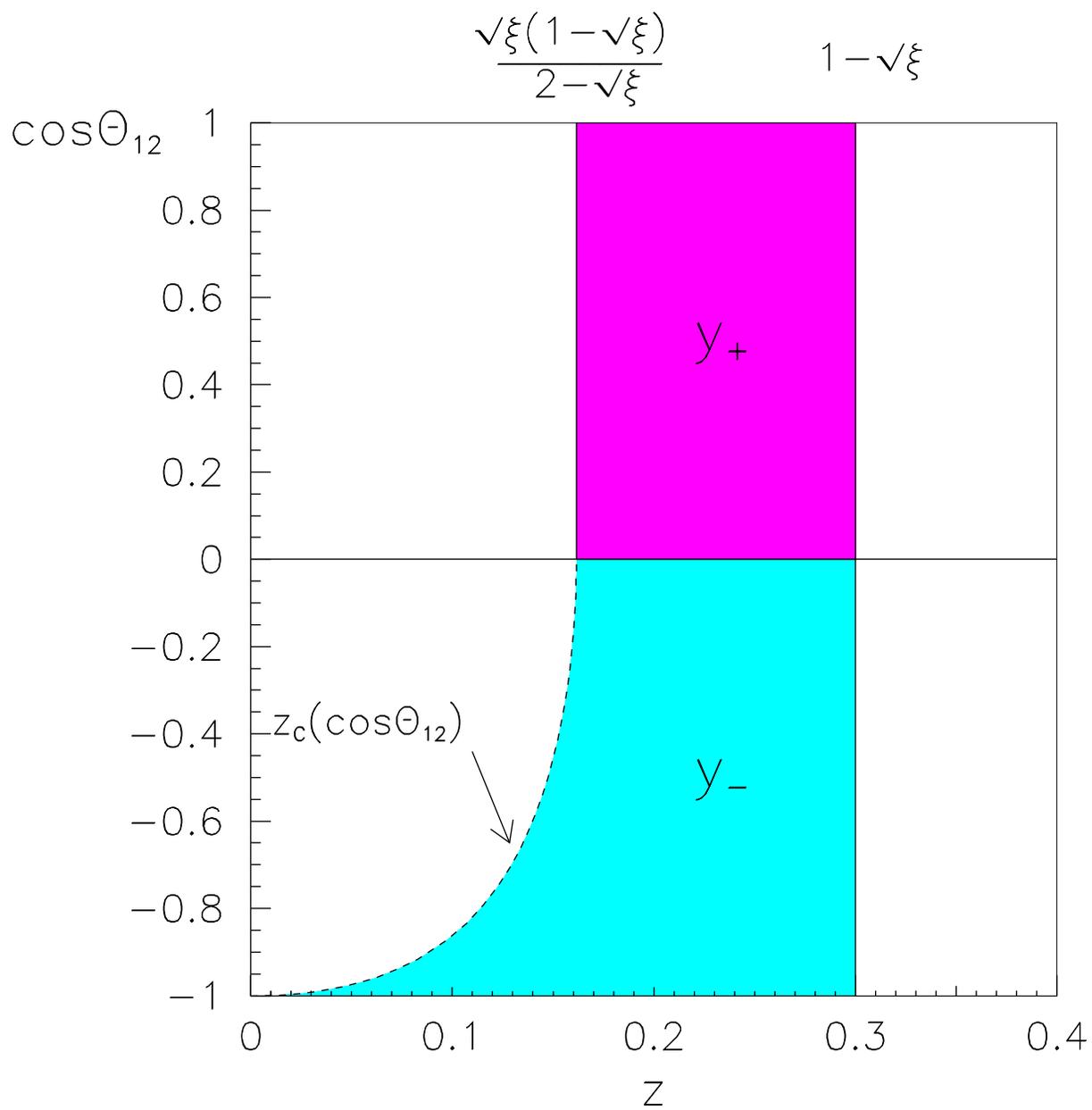
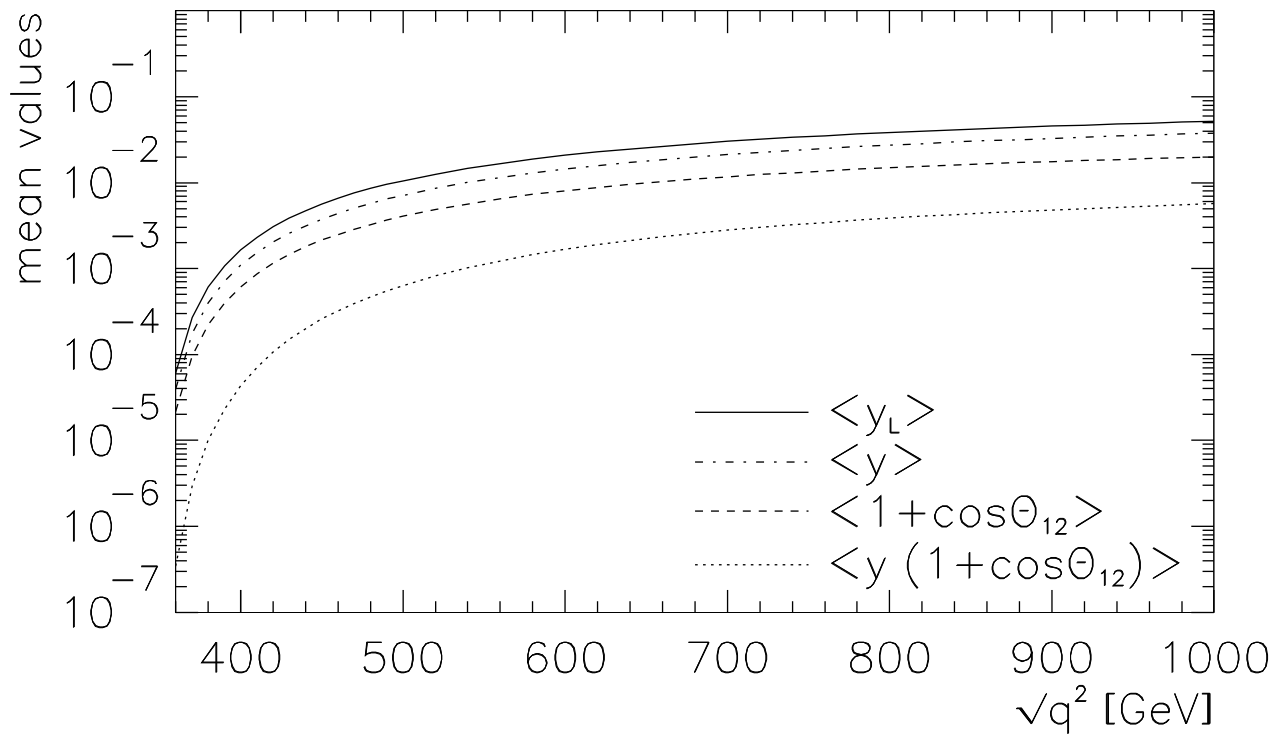
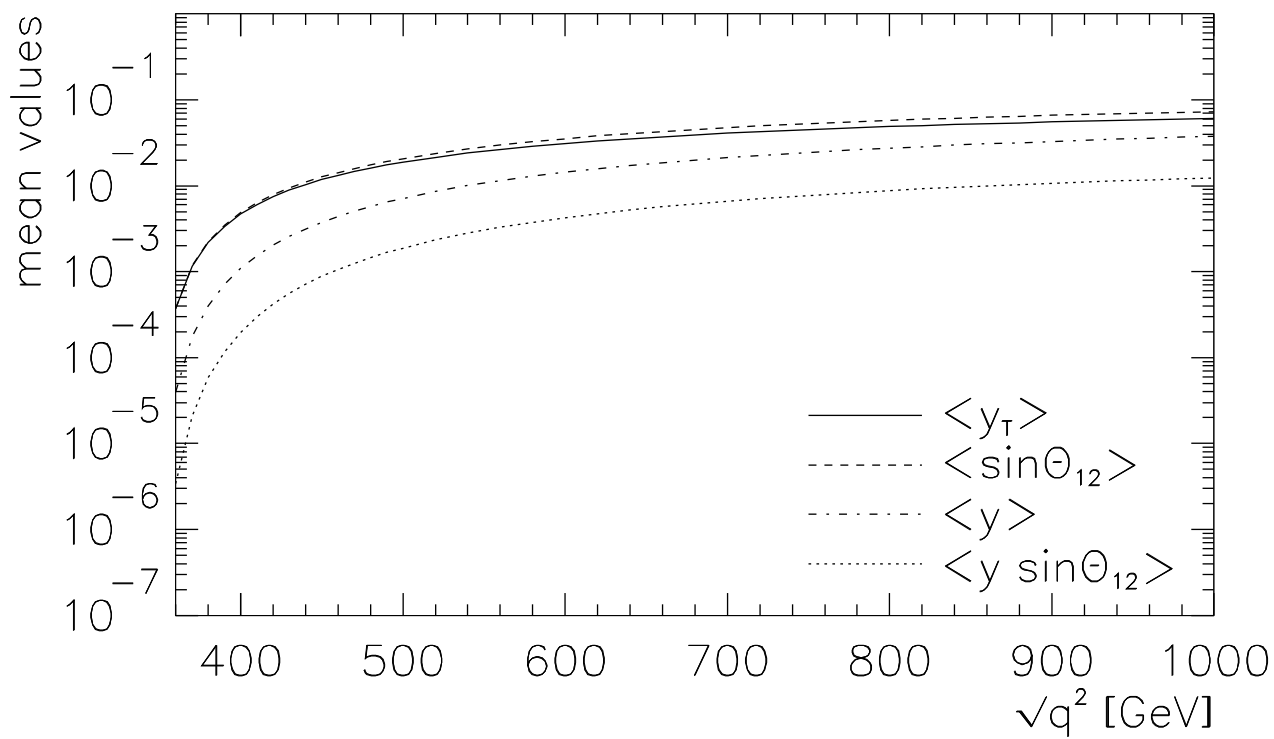
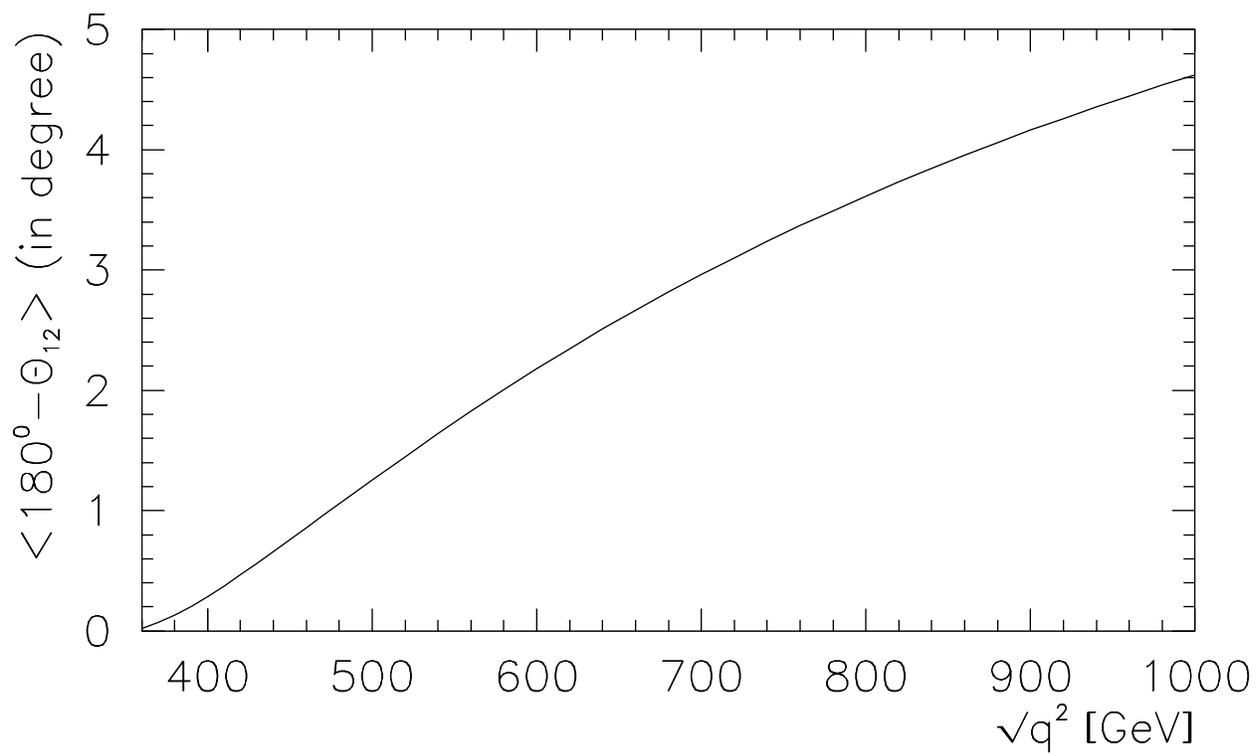
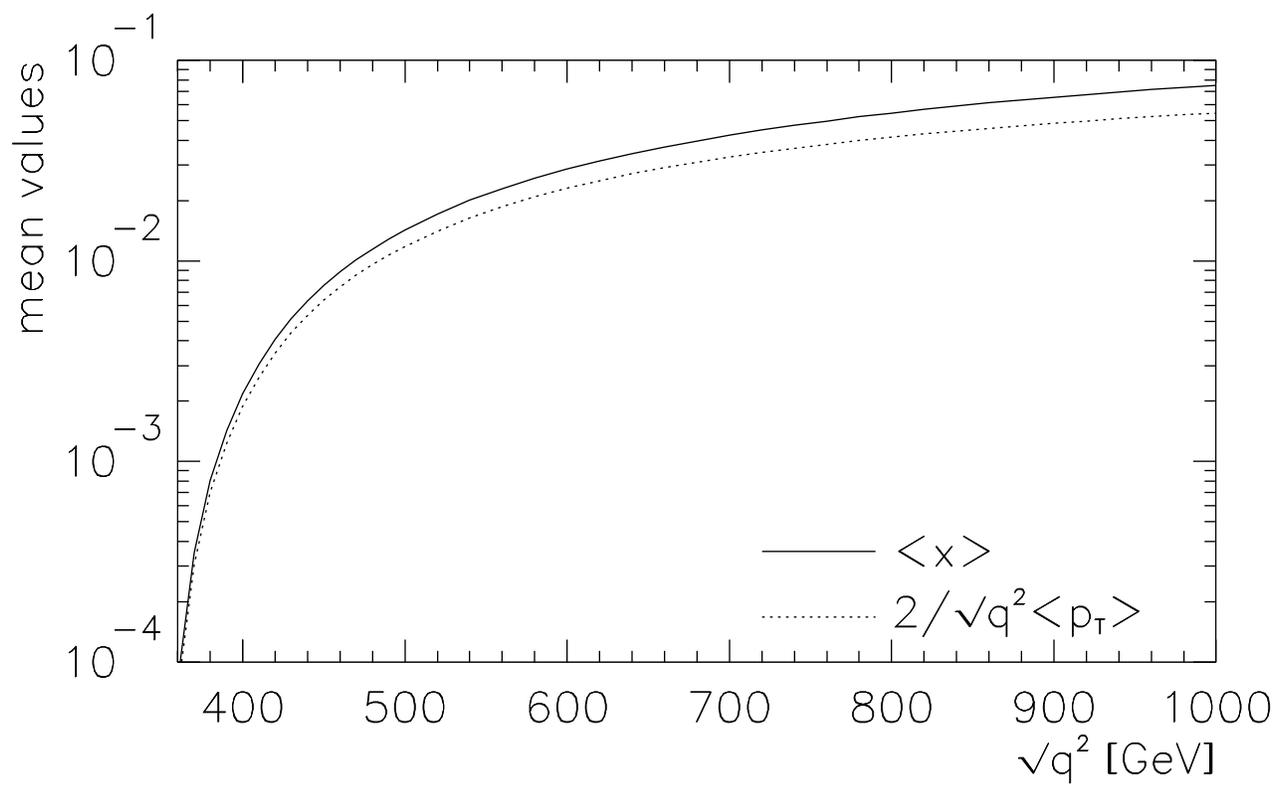
\end{document}